1    **Hybrid zone dynamics under weak Haldane's rule**

2    Ren-Xue Wang

3    BC Cancer Research Centre, BC Cancer Agency, Vancouver, British Columbia, Canada

5    *Running title: Wang, Hybrid zone dynamics under weak Haldane's rule*

7    *Correspondence:* Ren-Xue Wang, BC Cancer Research Centre, BC Cancer Agency,

8    675 West 10th Avenue, Vancouver, British Columbia, Canada V5Z 1L3. e-mail:

9    rwang@bccrc.ca. Phone: (604) 675 8000, extension: 7709. Fax: (604) 675 8185.

11      **The ability of genetic isolation to block gene flow plays a key role in the speciation of**

12      **sexually reproducing organisms. This paper analyses the hybrid zone dynamics affected**

13      **by "weak" Haldane's rule, namely the incomplete hybrids inferiority**

14      **(sterility/inviability) against the heterogametic (*XY* or *ZW*) sex caused by a Dobzhansky-**

15      **Muller incompatibility. Different strengths of incompatibility, dispersal and density-**

16      **dependent regulation are considered; and the gene flow and clinal structures of allele**

17      **frequencies in the presence of short-range dispersal (the stepping-stone model) are**

18      **examined. I show that a weak heterogametic hybrid incompatibility could constitute a**

19      **substantial barrier that could reduce gene flow and result in non-coincident and**

20      **discordant clines of alleles. It is found that the differential gene flow is more**

21      **pronounced under a stronger density-dependent regulation. This study provides a**

22      **mechanistic explanation for how an adaptive mutation, which may only have a marginal**

23      **fitness effect, could set a gene up as an evolutionary hot-spot.**

25      **Keywords:** Dobzhansky-Muller incompatibility, Gene flow, Haldane's rule, Hybrid

26      sterility, Hybrid inviability, Hybrid zone, Speciation.

28    Hybrids between closely related species of sexually reproduced organisms generally suffer

29    from poor fertility or viability. The question puzzled Darwin himself when formulating his

30    theory of evolution by natural selection was how evolution could maintain inheritable

31    changes to cause hybrid sterility or inviability. Several decades had passed since Bateson

32    (Bateson, 1909, Orr, 1996), Dobzhansky (Dobzhansky, 1937) and Muller (Muller, 1940,

33    Muller, 1942) articulated the evolutionary mechanism that produces such hybrid

34    incompatibility; it is through a process of the establishment of epistatic deficiencies between

35    diverging populations. Detrimental epistasis can be accumulated over time by genetic drifts

36    and/or adaptive mutations in separated populations (Wittbrodt et al., 1989, Ting et al., 1998,

37    Barbash et al., 2003, Presgraves et al., 2003, Brideau et al., 2006, Masly et al., 2006, Phadnis

38    & Orr, 2009) without facing negative selection or "adaptive valley" (Wright, 1988, Gavrilets,

39    1997). When hybridization occurs, such epistatic interaction could cause sterility or

40    inviability. For instance, from a diallelic ancestral population, an *aabb* genotype could evolve

41    into *AAbb* and *aaBB* genotypes respectively in two separate subpopulations, where the "*A*" or

42    "*B*" allele faces little or no negative selection for their establishment. If "*A*" and "*B*" alleles

43    are incompatible with each other and hybridization occurs, hybrids carrying *AaBb* genotype

44    will become inferior. This form of reproductive barriers is thus dubbed Dobzhansky-Muller

45    (DM) isolation.

46    Haldane's rule represents a major form of BDM isolation in early speciation (Coyne &

47    Orr, 1989, Wu & Davis, 1993, Laurie, 1997). Haldane's rule indicates that hybridization of

48    closely-related species almost always produces sterile or inviable $F_1$ offspring of the

49    heterogametic sex (male in *XY* sex determination and female in *ZW* determination) as

50    opposed to those of the homogametic sex if they are sex-biased. This rule is widely present in

51    almost all taxa investigated (Coyne & Orr, 1998), including mammals, fruit flies, butterflies,

52    birds and many other animals, and dioecious plants (Coyne & Orr, 1989, Wu & Davis, 1993,

53    Laurie, 1997). Heterogametic hybrid sterility/inviability appears to be a critical intermediate

54    step during speciation of many species. However, little is known about the role of a

55    heterogametic hybrid incompatibility in affecting population dynamics (Wang, 2003, Wang

56    & Zhao, 2008). Sex-biased BDM isolation is often the first appeared and a common (if not

57    the most common) form of reproductive isolation in many closely related species pairs, but

58    they have been rarely considered in the theoretical models of speciation with gene flow. The

59    existing theoretical studies on the divergence-with-gene-flow are mainly on the interplays

60    between prezygotic isolation (sexual selection) and ecological differentiations (Dieckmann &

61    Doebeli, 1999, Arnegard & Kondrashov, 2004, Gourbiere, 2004, Van Doorn, 2004, van

62    Doorn et al., 2009, also see Coyne & Orr, 2004, Fitzpatrick et al., 2008, Fitzpatrick et al.,

63    2009).

64       Hybrid zone analysis provides an excellent system for appreciating the role of Haldane's

65    rule in population dynamics during speciation (Wang & Zhao, 2008). Theoretical studies on

66    BDM isolation in affecting gene flow and population dynamics were first attempted by

67    Barton & Bengtsson (1986) and Gavrilets (1997). Using heuristic approximation, Bengtsson

68    and Barton (Barton & Bengtsson, 1986) developed a hybrid zone model with BDM isolation

69    and made a first attempt to analyse the neutral gene flow in such a scenario. It was concluded

70    that in the presence of autosome-linked BDM isolation, neutral gene flow could be reduced

71    greatly, but only when it was closely linked to one of the selected alleles. BDM isolation can

72    only constitute a strong barrier to neutral alleles in a hybrid zone when most of them are

73    closely linked to loci under selection (Barton & Bengtsson, 1986). By adapting a more

74    sophisticated fitness matrix, Gavrilets (Gavrilets, 1997) extended the analysis and concluded

75    that epistatic BDM isolation can build up a very strong barrier to neutral gene flow. Gavrilets

76    described in greater detail of the properties of hybrid zones under BDM selection, such as the

77    shapes of clines in relation to strength of selection, migration and linkage. Both of these

78    models (Barton & Bengtsson, 1986, Gavrilets, 1997) use numerical approximations when

79    analysing diallelic, autosome-linked epistatic selection. Sex chromosome-linked and sex

80    dependent selection was not considered.

81    Recently, Wang and Zhao (Wang & Zhao, 2008) developed a recursion model to analyse

82    hybrid zone structure and gene flow in a more sophisticated scenario – sex-dependent and

83    sex-chromosome-linked BDM isolation. Different from the earlier hybrid zone models under

84    BDM isolation (Barton & Bengtsson, 1986, Gavrilets, 1997), this model provides a more

85    accurate account of the effects of BDM isolation in hybrid zone dynamics. Wang and Zhao's

86    model (Wang & Zhao, 2008) separately considers the loci of $X$, $Y$, incompatible autosomes,

87    and neutral autosomes. The model allows the analysis of various alleles in a hybrid zone both

88    temporally and spatially. The analysis shows that in the presence of short-range dispersal

89    (stepping-stone model), a sex-biased hybrid incompatibility is an efficient barrier that

90    impedes gene flow across hybrid zones. Sex-biased hybrid incompatibilities differ from each

91    other depending on the sex that they select against, and the chromosomes where the

92    incompatible alleles are localized (Wang & Zhao, 2008).

93    The current paper focuses on the effects of "weak" Haldane's rule or the partial sex-biased

94    incompatibility that only causes inferiority of one sex in a fraction of carriers ranging from 0

95    to 100%. Between populations, weak sterility and inviability can arise as incidental

96    byproducts of divergence, such as adaptive mutations or genetic drifts, which may not be

97    readily observable in field and laboratory investigations. However, the population dynamics

98    under such otherwise unnoticeable partial isolation may be important for the formation of

99    Haldane's rule (Wang, 2003, Wang & Zhao, 2008). Many questions can be asked, such as,

100   how efficient a weak incompatibility could be as a genetic barrier; and to what extent it does

101   affect gene flow and hybrid zone structure; or how it is related to other factors in a hybrid

102   zone. The dynamics of such system may provide insights for understanding the evolutionary

103    mechanisms underlying speciation and Haldane's rule. Here, I quantitatively analyze the

104    influence of such partial incompatibility on the effective dispersal, density distribution, and

105    clinal structure of alleles across a hybrid zone. Variables that are considered also include

106    dispersal and density-dependent regulation.

107

108    **Definitions and assumptions**

109        Some common assumptions of population genetics are adopted. These include random

110    mating, discrete and non-overlapping generations, finite parental populations and passage of

111    alleles between generations at their probabilities (Endler, 1973, Slatkin, 1973). The parental

112    populations ($H_1$ and $H_2$ in the text, Wang & Zhao, 2008) are geographically separated and

113    genetically distinct.

114        *Sex-biased, two-locus BDM incompatibility.* Assume that between two diverging

115    populations, there is inferiority (sterility or inviability) that affects only hybrids of the

116    heterogametic sex. This is caused by detrimental epistatic interaction between two loci (two-

117    locus BDM incompatibility). The two incompatible alleles are localized on the sex

118    chromosome $X_1$ and an autosome $A_2$, which originate from different populations. These

119    alleles cause no fitness loss in their parental populations. The strength of an $X_1A_2$

120    incompatibility ($\psi$) ranges from 0 to 1,  namely that between 0 and 100% of the carriers of a

121    specific sex are inferior (sterile or inviable). I assume that hybrid inferiority is a continuous

122    quantitative trait – each incompatibility causes a fitness loss and produces fewer offspring in

123    average in accordance with its strength ($0 \leq \psi \leq 1$). The rest of the alleles are selectively

124    neutral and confer no fitness effects (loss or gain) in hybrids. In a deme, a fraction of hybrids

125    of affected sex are eliminated according to the strength of incompatibility. All other progeny

126    are equally fit, but subject to further segregation in later generations. For simplicity, we

127    assume male heterogamety (*XY*) but the results should apply to female heterogametic (*ZW*)

128    species as well.

129    *Density dependent regulation (r)*. In this study, density dependent regulation (*r*) is defined

130    as the ability of a population to adjust its size toward its carrying capacity (*N₀*). The effect of

131    a hybrid incompatibility is not density dependent (see below). However, the reduction of the

132    size caused by the incompatibility leads to a recovery of the deme toward its carrying

133    capacity.

134    *Gene flow and effective dispersal/migration*. Bengtsson (1985) has provided a clear

135    definition of effective dispersal/migration, "The effective migration rate, $m_e$, is that rate of

136    migration which would have the same evolutionary effect in a population with no genetic

137    barrier as the actual migration rate now has in the population with a barrier." For short-term

138    dispersal in the stepping stone model (Endler, 1973), dispersal only occurs between

139    neighbouring demes by a dispersal rate $\lambda$. I consider the rate of an allele entering the opposite

140    population across the zone as the "evolutionary effect (Bengtsson, 1985)". The effective

141    dispersal $\lambda_e$ is defined as the equivalent dispersal under no genetic isolation ($\psi = 0$) that could

142    achieve the same gene flow (the rate of an allele entering the opposite population) in a

143    scenario with a genetic incompatibility ($0 \leq \psi \leq 1$). In other words, the effective

144    dispersal/migration $\lambda_e$ is the product of $\lambda$ multiplied with the ratio of the rate of an allele

145    entering the opposite population in the presence of a genetic barrier (when $0 \leq \psi \leq 1$) to that

146    rate in the absence of a genetic barrier (when $\psi = 0$).

147    The meaning of the effective dispersal ($\lambda_e$) is the same as that the effective migration ($m_e$)

148    defined by (Bengtsson, 1985)(Barton & Bengtsson, 1986), and Gavrilets (Gavrilets, 1997).

149    However, Barton and Bengtsson (Barton & Bengtsson, 1986) and Gavrilets (Gavrilets, 1997)

150    only considered the flow of neutral markers. In this analysis, the flow of different alleles (*X*-

151    linked, *Y*-linked, incompatible autosomal, and neutral autosomal) are considered separately.

152    **Model**

153    The Wang and Zhao's recursion model for short-range dispersal (Wang & Zhao, 2008) is

154    extended for examining the influences of weak incompatibilities on hybrid zone dynamics. I

155    use the stepping-stone model for short-range dispersal (Endler, 1973, Wang & Zhao, 2008).

156    Briefly upon a scenario of secondary contact, a hybrid zone consisting of a chain of "*n*"

157    demes of equal size is formed by migration of both parental populations (*H₁* to *H₂*, see Wang

158    & Zhao, 2008); and the migration of mature adults occurs between adjacent demes. In every

159    generation, each deme loses $\lambda$ ($0 \leq \lambda \leq 0.5$) of its offspring to migration and at the same time

160    it receives input migration of $\lambda$ from the adjacent demes ($\lambda/2$ on each side). In the either end

161    of the chain, the deme exchanges migrants with the corresponding parental population. The

162    size of each parental population is 100 times that of the hybrid zone.

163    Four independently localized loci, each with two alleles are considered and denoted $X_1/X_2$,

164    $Y_1/Y_2$, $A_1/A_2$ and $C_1/C_2$. All these genotypes are expressed as $X_iX_jA_kA_lC_oC_p$ for females and

165    $X_iY_jA_kA_lC_oC_p$ for males, where *i*, *k*, and *o* represent the maternal origin alleles and *j*, *l*, and *p*

166    represent the paternal origin alleles. This diallelic system consists of $4^3 = 64$ possible

167    genotypic combinations of offspring for each sex (Wang & Zhao, 2008). Throughout the

168    paper, the subscripts 1 and 2 represent the population origin of alleles.

169    The frequency of each genotypic combination of either a sperm or an egg produced in

170    generation *t* is expressed as $p_{X_rA_sC_t}^{(t)sperm}$, $p_{Y_rA_sC_t}^{(t)sperm}$    and $p_{X_qA_sC_w}^{(t)egg}$, where *q, r, s, t, w,* or *z* represents the

171    population origin of the allele (1 or 2). Accordingly, the frequency of a given genotypic

172    combination in generation *t* in a hybridizing deme is the product of frequencies of a sperm

173    and an egg (Wang & Zhao, 2008) where:

174 $$u^{(t)} = p^{(t)}_{X_q Y_r A_s A_t C_w C_z} = p^{(t)egg}_{X_q A_s C_w} \bullet p^{(t)sperm}_{Y_r A_t C_z} \qquad (1)$$

175 or

176 $$v^{(t)} = p^{(t)}_{X_q X_r A_s A_t C_w C_z} = p^{(t)egg}_{X_q A_s C_w} \bullet p^{(t)sperm}_{X_r A_t C_z} \qquad (2)$$

177 The $u^{(t)}$ or $v^{(t)}$ represents the frequency of a genotype $X_q Y_r A_s A_t C_w C_z$ or $X_q X_r A_s A_t C_w C_z$ in males

178 or females in generation $t$ before migration respectively.

179     The frequencies of genotypic combinations for each sex in a chain of demes in generation

180 $t$ are expressed by ($n \times$ m) matrices ($m = 64$), $U_0^{(t)}$ (for male) and $V_0^{(t)}$ (for female).

181 $$U_0^{(t)} = \begin{bmatrix} u^{(t)}_{11} & \cdots & u^{(t)}_{1m} \\ \vdots & & \vdots \\ \vdots & & \vdots \\ u^{(t)}_{n1} & \cdots & u^{(t)}_{nm} \end{bmatrix} \text{ and } V_0^{(t)} = \begin{bmatrix} v^{(t)}_{11} & \cdots & v^{(t)}_{1m} \\ \vdots & & \vdots \\ \vdots & & \vdots \\ v^{(t)}_{n1} & \cdots & v^{(t)}_{nm} \end{bmatrix} \qquad (3)$$

182 in which each row vector represents the frequencies of genotype combinations in deme $i$

183 computed by (1) or (2). The range of $i$ is between 1 to $n$ corresponding to the positions of

184 demes and the range of $j$ is between 1 to $m$ ($m = 64$ in this four loci scenario) corresponding

185 to genotypic combinations.

186     The size of a compatible genotypic combination in deme $i$ is thus:

187 $$w^{(t)}_{ij} = N^{(t-1)}_i u^{(t)}_{ij} \text{ and } z^{(t)}_{ij} = N^{(t-1)}_i v^{(t)}_{ij} \qquad (4)$$

188 The size of an incompatible genotypic combination in deme $i$ is:

189 $$w^{(t)}_{ij} = (1-\psi) N^{(t-1)}_i u^{(t)}_{ij} \text{ and } z^{(t)}_{ij} = (1-\psi) N^{(t-1)}_i v^{(t)}_{ij} \qquad (5)$$

190 Here, $\psi$ ($0 \leq \psi \leq 1$) is the strength of an incompatibility. A $\psi$ fraction of an incompatible

191 genotype will be inferior and eliminated in the deme; ($1 - \psi$) will survive.

192     The density-dependent regulation of the population/deme size is based on the classic

193 logistic model of population growth (see Hartl & Clark, 1997, page 31) and its strength is

194     expressed as $r$ (the intrinsic rate of increase). When the male is the affected sex, the size of

195     deme $i$ in generation $t$ before migration is calculated by:

196

$$N_i^{(t)} = \sum_j^m w_{ij}^{(t)} + r\sum_j^m w_{ij}^{(t)} \left(1 - \frac{\sum_j^m w_{ij}^{(t)}}{N_0}\right) \qquad (6)$$

197     When the female is the affected sex, the size of deme $i$ is:

198

$$N_i^{(t)} = \sum_j^m z_{ij}^{(t)} + r\sum_j^m z_{ij}^{(t)} \left(1 - \frac{\sum_j^m z_{ij}^{(t)}}{N_0}\right) \qquad (7)$$

199     Here, $N_0$ is the carrying capacity or the optimal population size. In each generation before

200     migration, the size of a population will be regulated by $r$ until the balance is reached to satisfy

201     $N^{(t)} = N^{(t-1)} + rN^{(t-1)}\left(1 - \frac{N^{(t-1)}}{N_0}\right)$ (Hartl & Clark, 1997). The $r$ and $\psi$ together determine the

202     deme size.

203     The relative density of a deme before migration is:

204

$$\rho = N_n^{(t)} / N_0 \qquad (8)$$

205     Here, $\rho$ is $0 \le \rho \le 1$ and $N_0$ is the carrying capacity, which is the same for all demes here.

206     The proportional sizes of genotypic combinations that will remain in the original demes

207     after elimination of incompatible offspring are:

208

$$W^{(t)} = (1-\lambda)\begin{bmatrix} w_{11}^{(t)} & \cdots & w_{1m}^{(t)} \\ \vdots & & \vdots \\ \vdots & & \vdots \\ w_{n1}^{(t)} & \cdots & w_{nm}^{(t)} \end{bmatrix} \quad \text{and} \quad Z^{(t)} = (1-\lambda)\begin{bmatrix} z_{11}^{(t)} & \cdots & z_{1m}^{(t)} \\ \vdots & & \vdots \\ \vdots & & \vdots \\ z_{n1}^{(t)} & \cdots & z_{nm}^{(t)} \end{bmatrix} \qquad (9)$$

209     The proportional sizes of genotypic combinations that will be migrating in the direction from

210     $H_1$ to $H_2$ are:

211 $$W_{mig1}^{(t)} = \frac{\lambda}{2} \begin{bmatrix} N_0 & 0 & \cdots & 0 \\ w_{11}^{(t)} & w_{12}^{(t)} & \cdots & w_{1m}^{(t)} \\ \vdots & & & \vdots \\ w_{(n-1)1}^{(t)} & w_{(n-1)2}^{(t)} & \cdots & w_{(n-1)m}^{(t)} \end{bmatrix} \quad \text{and} \quad Z_{mig1}^{(t)} = \frac{\lambda}{2} \begin{bmatrix} N_0 & 0 & \cdots & 0 \\ z_{11}^{(t)} & z_{12}^{(t)} & \cdots & z_{1m}^{(t)} \\ \vdots & & & \vdots \\ z_{(n-1)1}^{(t)} & z_{(n-1)2}^{(t)} & \cdots & z_{(n-1)m}^{(t)} \end{bmatrix} \quad (10)$$

212 Here, the first row represents the inputs from $H_1$ with the genotypes $X_1X_1A_1C_1C_1$ for the

213 female and $X_1Y_1A_1C_1C_1$ for the male. The sum of frequencies of either sex equals to 1. We

214 assume that the carrying capacity ($N_0$) is the maximum deme size. The proportional sizes of

215 genotypic combinations that will be migrating in the direction from $H_2$ to $H_1$ are:

216 $$W_{mig2}^{(t)} = \frac{\lambda}{2} \begin{bmatrix} w_{21}^{(t)} & \cdots & w_{2(m-1)}^{(t)} & w_{2m}^{(t)} \\ \vdots & & & \vdots \\ w_{n1}^{(t)} & \cdots & w_{n(m-1)}^{(t)} & w_{nm}^{(t)} \\ 0 & \cdots & 0 & N_0 \end{bmatrix} \quad \text{and} \quad Z_{mig2}^{(t)} = \frac{\lambda}{2} \begin{bmatrix} z_{21}^{(t)} & \cdots & z_{2(m-1)}^{(t)} & z_{2m}^{(t)} \\ \vdots & & & \vdots \\ z_{n1}^{(t)} & \cdots & z_{n(m-1)}^{(t)} & z_{nm}^{(t)} \\ 0 & \cdots & 0 & N_0 \end{bmatrix} \quad (11)$$

217 Here, the last row represents the inputs from $H_2$ with the genotypes $X_2X_2A_2C_2C_2$ *for* the

218 female and $X_2Y_2A_2C_2C_2$ for the male.

219 Therefore, the proportional sizes of genotypic combinations in the zone after migration

220 are:

221 $$M_{mig}^{(t)} = W^{(t)} + W_{mig1}^{(t)} + W_{mig2}^{(t)} \quad \text{and} \quad F_{mig}^{(t)} = Z^{(t)} + Z_{mig1}^{(t)} + Z_{mig2}^{(t)} \quad (12)$$

222 Thus, the frequency of a genotype combination after migration will be:

223 $$\omega_{ij}^{(t)} = \frac{(1-\lambda)w_{ij}^{(t)} + \frac{\lambda}{2}\left(w_{(i+1)j}^{(t)} + w_{(i-1)j}^{(t)}\right)}{(1-\lambda)\sum_j^m w_{ij}^{(t)} + \frac{\lambda}{2}\sum_j^m\left(w_{(i+1)j}^{(t)} + w_{(i-1)j}^{(t)}\right)} \quad (13)$$

224 and the frequency of a female genotypic combination will be:

225 $$\kappa_{ij}^{(t)} = \frac{(1-\lambda)z_{ij}^{(t)} + \frac{\lambda}{2}\left(z_{(i+1)j}^{(t)} + z_{(i-1)j}^{(t)}\right)}{(1-\lambda)\sum_j^m z_{ij}^{(t)} + \frac{\lambda}{2}\sum_j^m\left(z_{(i+1)j}^{(t)} + z_{(i-1)j}^{(t)}\right)} \quad (14)$$

226 The simulations and graphics used in the figures were generated with MATLAB

227     (Code available by request).

228

229  **Results**

230  *Gene flow under a partial sex-biased hybrid incompatibility – effective dispersal*

231  The effective dispersal ($\lambda_e$) represents the strength of gene flow or introgression of an

232  allele. The $\lambda_e$ of alleles on different chromosomes are different as shown in Figure 1. Figure 1

233  shows the distribution of $\lambda_e$ under the influence of an $X_1A_2$ incompatibility ranging from 0 to

234  1 ($0 \leq \psi \leq 1$, $X$-axis) that is in a scenario of a mild density-dependent regulation ($r = 0.05$

235  and $\lambda = 0.1$ (Figure 1A), 0.2 (Figure 1B) and 0.4 (Figure 1C). It is evident that the

236  introgressions of the incompatible loci ($X$ and $A$) are highly asymmetrical under a weak

237  Haldane's rule. For instance, in the scenario of $\lambda = 0.4$ (Figure 1C), a 10% ($\psi = 0.1$) $X_1A_2$

238  incompatibility results in a $\lambda_e$ of $X_1$ and $A_2$ about 0.1060 and 0.158 respectively, and a $\lambda_e$ of $X_2$

239  and $A_1$ of 0.440 and 0.366 respectively at the "equilibrated point" (2000 generations). The

240  non-incompatible alleles ($X_2$ and $A_1$) have a much higher introgression comparing to their

241  incompatible counterparts ($X_1$ and $A_2$). As shown in Supplementary Figure 1, a higher

242  density-dependent regulation extends the differentiation of introgression between the $X$ and $A$

243  loci. In a scenario when $r = 0$, $\lambda = 0.4$ and $\psi = 0.2$, the $\lambda_e$ of $X_1$ and $X_2$ would be 0.066 and

244  0.220 respectively. However, when $r$ increases to 0.1 in the same scenario, the $\lambda_e$ of $X_1$ and $X_2$

245  would become 0.118 and 0.51 respectively (Supplementary Figure 1A & 1C). Also, the

246  compatible counterparts of the incompatible alleles have a higher tendency to introgress

247  comparing to the alleles at neutral loci when density-dependent regulation is strong. The $\lambda_e$ of

248  these compatible counterparts can be much higher than the actual $\lambda$, but $\lambda_e$ of a neutral allele

249  is always lower than $\lambda$ (Figure 1 and Supplementary Figure 1). A strong flow of non-

250  incompatible alleles ($X_2$ and $A_1$ in this example) would constitute a strong homogenization

251  force causing the collapse of the genetic barrier. Furthermore, a density-dependent regulation

252  could substantially are negatively correlated with the barrier strength against the neutral flow.

253  As shown in Supplementary Figure 1D, under a relatively strong density-dependent

254    regulation ($r = 0.5$), the introgression of a neutral allele is very close to the scenario with no

255    incompatibility (the reference lines in Supplementary Figure 1D). Interestingly, the trends of

256    $\lambda_e$ of all alleles plateau long before a full strength BDM incompatibility ($\psi = 1$), suggesting

257    that even a milder $X_1A_2$ incompatibility, as low as 20% ($\psi = 0.2$), could mount significant

258    impedance on gene flow that is almost comparable to a 100% incompatibility ($\psi = 1$).

259

260    *The effective of partial sex-biased incompatibilities on the cline structures of various*

261    *alleles*

262    Genetic isolation could cause clinal non-coincidence and discordance of alleles in a hybrid

263    zone. I examined the effects of a weak $X_1A_2$ unidirectional incompatibility on the clinal

264    structure in relation to variations of $\psi$, $\lambda$, and $r$. Figure 2 shows some clines of alleles in a

265    number of representative scenarios. It can be seen that an $X_1A_2$ unidirectional incompatibility

266    causes significant clinal non-coincidence and discordance. The cline of a neutral locus is

267    wider, less steep ($P_{C1}$ in Figure 2), and almost symmetrically distributed, but the clines of

268    incompatible loci are steeper, narrower and asymmetrically distributed ($P_{X1}$ and $P_{A1}$ in Figure

269    2). Figure 2A, B and C show the clines caused by a 10% $X_1A_2$ incompatibility ($\psi = 0.1$). The

270    non-coincidence of clines is rather obvious and significant under the 10% incompatibility but

271    to a lesser extent in comparison with those under a full scale $X_1A_2$ heterogametic

272    incompatibility ($\psi = 1$), which is  shown in Figure 2D, E and F. Figure 2 also shows that a

273    stronger density-dependent regulation produces less steep clines of the $Y$ and neutral $C$ loci,

274    similar to those formed through neutral diffusion (Endler, 1973) (Endler, 1977, Barton &

275    Gale, 1993). The clines of $X$ and $A$ are more asymmetrically distributed, in which the $X$ locus

276    shows the most asymmetry. This suggests that a very weak $X_1A_2$ incompatibility under a

277    stronger density-dependent regulation could lead to a pronounced asymmetric introgression

278    of $X$ and $A$ loci, but not in the case of $Y$ and $C$ loci. Among all loci, the cline of the $X$ locus is

279    the narrowest and steepest, which could likely facilitate faster divergences of *X*-linked loci

280    during speciation.

281

282    *The hybrid sink effects of weak incompatibilities*

283    To better appreciate the fitness effect of a weak $X_1A_2$ heterogametic incompatibility, the

284    density depression in a hybrid zone as a consequence of elimination of inferior hybrids was

285    examined. The zone here acts as a "sink" because the production of inferior hybrids results in

286    an inward net flow of migrants towards the centre of the zone (the hybrid sink effect)

287    (Barton, 1980, Barton & Bengtsson, 1986). Distribution of relative density ($0 \leq \rho \leq 1$) across

288    a hybrid zone is the function of the incompatibility strength ($\psi$), dispersal rate ($\lambda$), and

289    density-dependent regulation (*r*). Figure 3 shows the distribution of relative density across a

290    hybrid zone in different scenarios at equilibrium. The density distribution in the zone displays

291    an asymmetrical, "V" shape depression.

292    In this 10 deme scenario, if density-dependent regulation is absent (*r* = 0), the lowest

293    depression point is at deme 5; if density-dependent regulation is present (*r* = 0.05 or 0.2),

294    lowest depression point shifts to further left to deme 4. This suggests that the asymmetrical

295    pressure increases when density-dependent regulation is larger. The density depression is

296    more pronounced when the strength of the incompatibility is higher, density-dependent

297    regulation is lower, or dispersal rate is ($\lambda$) is higher (Figure 3). There is a good correlation

298    between the density distribution and the percentages of inferior offspring produced in a

299    hybrid zone in each generation at equilibrium (See Supplementary Figure 2).

300    Next, I examined the relationship between the density depression and the strength of

301    density-dependent regulation (*X*-axis). Figure 4 shows the equilibrated density of Deme No.5

302    in response to the varying *r* values with a 5, 10, 30 or 100% incompatibility strength ($\psi$ =

303    0.05, 0.1, 0.3 or 1) and 10 or 40% dispersal ($\lambda$ = 0.1 or 0.4). A strong density-dependent

304    regulation can almost entirely compensate for the depression caused by an $X_1A_2$

305    incompatibility (Figure 3 and Figure 4). The analysis here also suggests that when density-

306    dependent regulation is relatively weak, a weak $X_1A_2$ heterogametic incompatibility ($\psi \ll 1$)

307    could still be a rather efficient barrier and causes a substantial density depression (Figure

308    s3A, 3B and 4). However, if density-dependent regulation is strong and the $r$ value is large ($r$

309    $> 0.2$), even a full strength heterogametic incompatibility ($\psi = 1$) would only cause an

310    insignificant density depression (Figure 4). Density-dependent regulation is an important

311    factor in regulating the density and the size of demes in a hybrid zone.

312    **Discussion**

313        The current paper is an attempt to examine the role of a weak BDM incompatibility in a

314    hybrid zone system. It provides a more quantitative account of how "slightly lessened fertility

315    (see 'Charles Darwin's letters: a selection, 1825-1859'Darwin & Burkhardt, 1998)" caused by

316    weak sex-biased inferiority (*i.e.* weak Haldane's rule) could contribute to the genetic

317    divergence and potentially the establishment of complete reproductive isolation during

318    speciation. The gene frequency changes and gene flow between diverging populations during

319    speciation are at the heart of population genetics. I found that a weak sex-biased hybrid

320    incompatibility can profoundly affect gene flow past a hybrid zone. A weak, sex-biased BDM

321    incompatibility provides *a)* a significant reinforcement pressure to drive further divergence of

322    population; and *b)* a substantial drive of the asymmetrical flow of incompatible loci (Figures

323    1), which leads to significant clinal non-coincidence and discordance (Figure 2).

324        The impedance of the flow of incompatible genes across a hybrid zone is no surprising.

325    Such gene flow usually results in abrupt clines (Barton & Gale, 1993). It is interesting,

326    however, that such impedance can be achieved through a weak, sex-biased BDM

327    incompatibility. As shown in Figure 1, a weak incompatibility with 10-20% ($\psi = 0.1$-$0.2$)

328    strength is sufficient to confer a significant reduction of gene flow across a hybrid zone of an

329    incompatible allele. The effective dispersal of most alleles under a 20% incompatibility is

330    close to that of a full strength sex-biased incompatibility (Figure 1). Such reduced gene flow

331    results in characteristic clinal structures in the hybrid zone. The width, steepness and spatial

332    distribution of allele clines (Figure 2) are similar, but to lesser extent, to those under a full

333    strength incompatibility (Wang & Zhao, 2008). Furthermore, a weak compatibility can also

334    cause a substantial density depression in the hybrid zone that is only moderately shallower

335    than a full strength sex-biased incompatibility (Figure 3). In essence, a hybrid zone in the

336    presence of weak sex-biased BDM isolation would act like a filter: it selectively blocks the

337    exchange of incompatible ones (e.g. $X_1$ and $A_2$ in an $X_1A_2$ incompatibility) and at the same

338    time, allows almost free exchange of neutral loci similar to that through neutral diffusion ($P_{C1}$

339    in Supplementary Figure 1C and 1D). The introgression of non-incompatible alleles, such as

340    the $X_2$ and $A_1$ in the case of $X_1A_2$ incompatibility, would be higher than that of neutral alleles

341    (Supplementary Figure 1C and D). The most affected loci by a weak Haldane's rule are on

342    the $X$ chromosomes. These trends are more pronounced when density-dependent regulation is

343    high.

344      The real question is how relevant these trends are to genetic divergence and speciation. It

345    is conceivable that a hybrid zone (tension zone) with a sex-biased isolation is highly unstable

346    and mobile, because of asymmetric gene flow and incomplete isolation. Maintaining such a

347    hybrid zone would have to involve other selection forces. For instance, a hybrid zone could

348    be stabilized if alleles related to BDM isolation have selective advantages in their own

349    respective localities; or when it coincides with density troughs and/or habitat boundaries with

350    environmental differentiations, which is indeed often the case in nature (Hewitt, 1988, Rice &

351    Hostert, 1993). The population dynamics described in this study may be indicative of how

352    sex-biased BDM incompatibilities is established and preserved during speciation.

353      Studies on diverse taxa have shown that natural selection caused by habitat shifts and

354    environmental changes can lead to extremely rapid genetic divergence and ecological

355    segregation, such as soapberry bugs (Carroll, 1997 #716;Carroll, 2003 #727), threespine

356    sticklebacks (Schluter, 1996, Albert et al., 2008, Berner et al., 2009), cichlids (Barluenga et

357    al., 2006), Neotropical guppies (Reznick et al., 1997), island lizards (Losos et al., 1997), and

358    rainforest passerines (Smith et al., 1997). For instance, after the introduction of an alternative

359    host into North America in less than 100 generations, the "derived-type" soapberry bugs

360    (*Jadera hematoloma*) had evolved epistatic incompatibilities from the "ancestral-type"

361    (Carroll et al., 2003). The "derived-type" and "ancestral-type" were genetically diverged in

362    the feeding morphology, growth rate, survival and fecundity (Carroll et al., 1997). In

363    threespine sticklebacks, the analysis of neutral microsatellites markers indicate that adaptive

364    divergence and partial reproductive isolation often coexist between two parapatric

365    populations (Berner et al., 2009). In these fish, some genomic regions, including the sex-

366    determining chromosome region, have the largest effect on adaptive traits by QTL analysis,

367    suggesting a rapid fixation of adaptive mutations and uneven divergence of different

368    chromosomal regions (Albert et al., 2008); ecological selection also drove rapid evolution of

369    extrinsic postzygotic isolation (Gow et al., 2007) and prezygotic isolation (Boughman et al.,

370    2005) between benthic and limnetic sticklebacks.

371        In recent years, many hotspot genes and mutations have been discovered (Stern &

372    Orgogozo, 2009). The evolutionary changes in the hotspots are often associated with specific

373    adaption (ffrench-Constant et al., 1998, Wichman et al., 1999, Shindo et al., 2005); and/or

374    with sex-biased incompatibility and Haldane's rule in various species pairs (Wittbrodt et al.,

375    1989, Ting et al., 1998, Barbash et al., 2003, Presgraves et al., 2003, Brideau et al., 2006,

376    Masly et al., 2006, Mihola et al., 2009, Phadnis & Orr, 2009, Tang & Presgraves, 2009).

377    These studies support the notion that the rapid evolution of isolating mechanisms is not

378    merely the by-products of genetic drifts or neutral divergence, as many once believed, in

379    environments with high ecological differentiations in sympatry or parapatry (Carroll et al.,

380    1997, Reznick et al., 1997, Orr & Smith, 1998, Wang, 2003, Wang & Zhao, 2008).

381        The interplays between adaptation and reproductive isolation hold the key for the

382    evolution of sexually reproducing organisms (Butlin et al., 2008, Fitzpatrick et al., 2008). In

383    light of the current analysis, one can speculate that in a highly differentiated environment, the

384    adaptive changes in subpopulations would more likely be preserved during speciation, if they

385    happen to cause hybrid inferiority. This mechanism may have led to the dominance of BDM

386    isolation and Haldane's rule. This population dynamics provides a glimpse of how a

387   speciation gene comes into being at an early stage of population divergence. In a

388   megapopulation consisting of many clades with complex environmental differentiations, an

389   adaptive mutation could set the gene up as a hotspot that initiates a runaway evolutionary

390   process toward more complete reproductive isolation. With such dynamics, local adaptation

391   and postzygotic isolation could promote each other during genetic divergence and population

392   differentiation. This model may also be extended to test some controversial issues in

393   evolutionary biology. For instance, whether or not secondary contact is a prerequisite for the

394   formation of hybrid zone and species; whether or not hybrid zones are the sites of

395   'reinforcement' (Dobzhansky, 1940) – the evolution of prezygotic barriers to gene exchange

396   in response to selection against hybrids (Harrison, 1993). These are the fundamental

397   questions in the field of population genetics that remain controversial for many years

398   (Paterson, 1978, 1982; Butlin, 1987, 1989).

399

400

401   **Acknowledgments**

402   I thank Tania Kastelic and Philip Tang for help with English editing.

403     **Figure 1.** The effective dispersal ($\lambda_e$) of different alleles in a scenario of $r = 0.05$ with

404     dispersal of (**A**) $\lambda = 0.1$; (**B**) $\lambda = 0.2$; and (**C**) $\lambda = 0.4$. The strength of an $X_1A_2$ heterogametic

405     incompatibility varies from 0 to 1 (*X*-axis). The horizontal dotted line in each panel is the

406     reference line, which represents a scenario of no genetic incompatibility between two

407     parental populations.

408

409     **Figure 2.** The non-coincident and discordant clines of the representative loci, *X*, *Y*, *A* and

410     *C* across a hybrid zone consisting of 10 demes with different density-dependent regulations

411     (*r*) and barrier strengths ($\psi$ – caused by an $X_1A_2$ heterogametic incompatibility) at 40%

412     dispersal ($\lambda = 0.4$). (**A**) $r = 0$, $\psi = 0.1$; (**B**) $r = 0.05$, $\psi = 0.1$; (**C**) $r = 0.4$, $\psi = 0.1$; (**D**) $r = 0$, $\psi$

413     = 1; (**E**) $r = 0.05$, $\psi = 1$; (**F**) $r = 0.4$, $\psi = 1$.

414

415     **Figure 3.** The relative density depression in an equilibrated hybrid zone (*t* = 2000)

416     consisting of 10 demes with the various strengths ($\psi$) of an $X_1A_2$ heterogametic

417     incompatibility and different density-dependent regulation, (**A** and **D**. $r = 0$; **B** and **E**. $r =$

418     0.05; and **C** and **F**. $r = 0.2$). In each panel, the alternate lines from top to bottom represent the

419     strength of an $X_1A_2$ heterogametic incompatibility of $\psi = 0.05, 0.1, 0.2, 0.3, 0.5$ or 1.

420

421     **Figure 4.** The equilibrated density distribution in Deme No.5 under different density-

422     dependent regulation in a scenario of 10 demes with different strengths of an $X_1A_2$

423     heterogametic incompatibility ($\psi = 0.05, 0.1, 0.3$ and 1) and dispersal (**A**) $\lambda = 0.1$ (**B**) $\lambda = 0.4$.

424

425     **Figure 5.** The population density over time (generations) in the Deme No.5 with different

426     strength of BDM incompatibility ($\psi$). From top to bottom, the alternate lines from top to

427     bottom represent a $\psi$ value of 0.05, 0.1, 0.2, 0.3, 0.5 or 1, respectively.

428

429    **Supplementary Figure 1.** The effective dispersal ($\lambda_e$) of different alleles in scenarios of

430    (**A**) $r = 0$, (**B**) $r = 0.05$, (**C**) $r = 0.1$, and (**D**) $r = 0.5$ with 40% dispersal ($\lambda = 0.4$) and the

431    strength ($\psi$) of an $X_1A_2$ heterogametic incompatibility varying from 0 to 1 ($X$-axis). The

432    horizontal dotted line in each panel is the reference line representing a scenario of no

433    incompatibility, in which there is no genetic isolation two parental populations.

434

435    **Supplementary Figure 2.** The percentage of inferior offspring under a complete $X_1A_2$

436    heterogametic incompatibility ($\psi = 1$) in each generation at equilibrium. The variations of

437    parameters are: $r = 0$, 0.05, and 0.2; $\lambda = 0.05$, 0.2 and 0.4.

438

439    **Supplementary Figure 3.** The time (generations) required for establishing equilibrium

440    under a weak $X_1A_2$ heterogametic incompatibility. The cutoff of equilibration is set at the 1%

441    collective size of all 10 demes. The hybrid zone with a size change smaller than the cutoff

442    between generations is considered at equilibrium. Here, the strength of the incompatibility

443    ($\psi$) varies from 0 to 1 ($X$-axis). Dispersal of 10, 20 and 40% ($\lambda = 0.1$, 0.2 and 0.4) are

444    considered. (**A**) $r = 0$; (**B**) $r = 0.05$; (**B**) $r = 0.1$; and (**C**) $r = 0.5$.

445

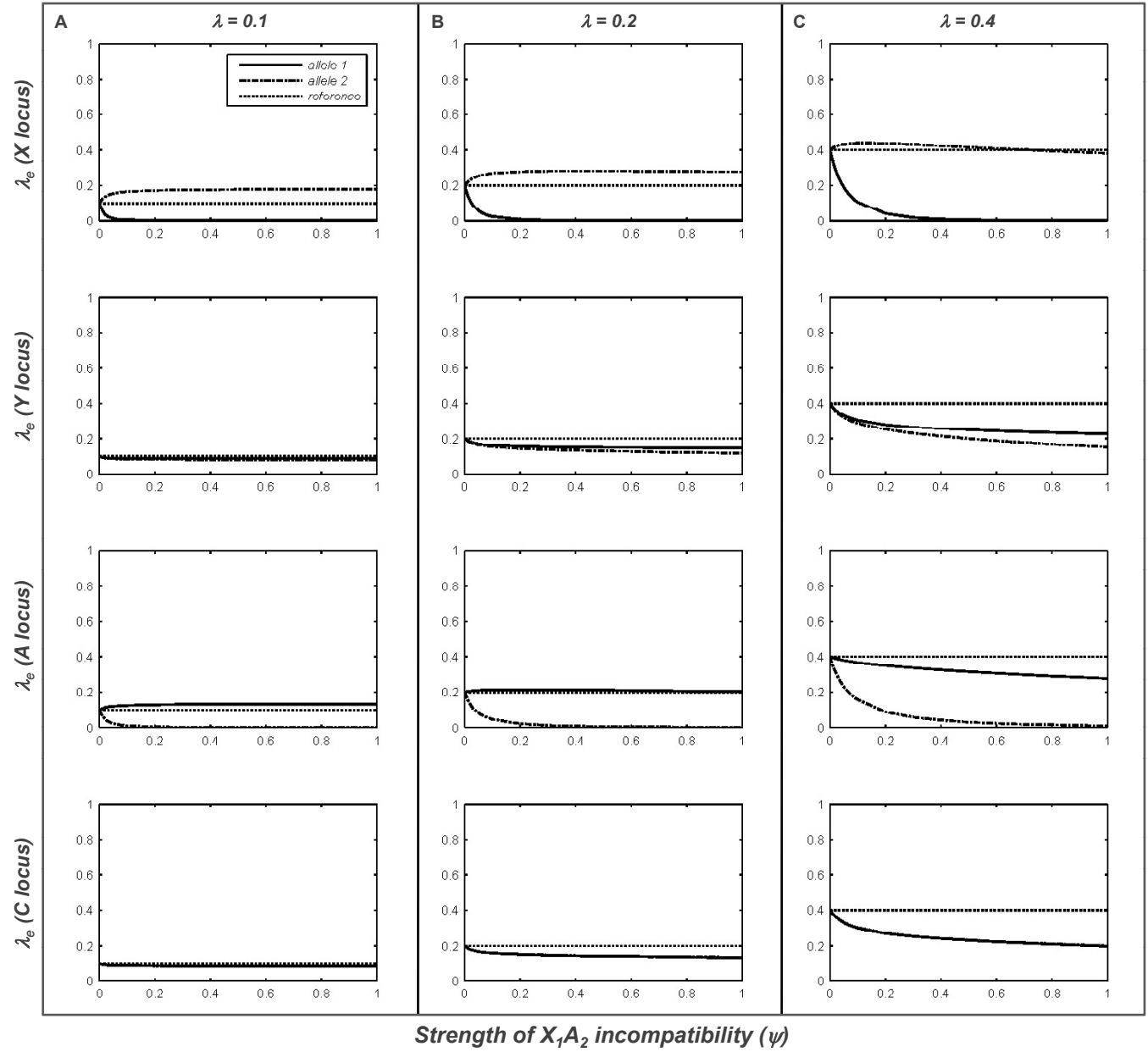

**Strength of X₁A₂ incompatibility (ψ)**

Figure 2

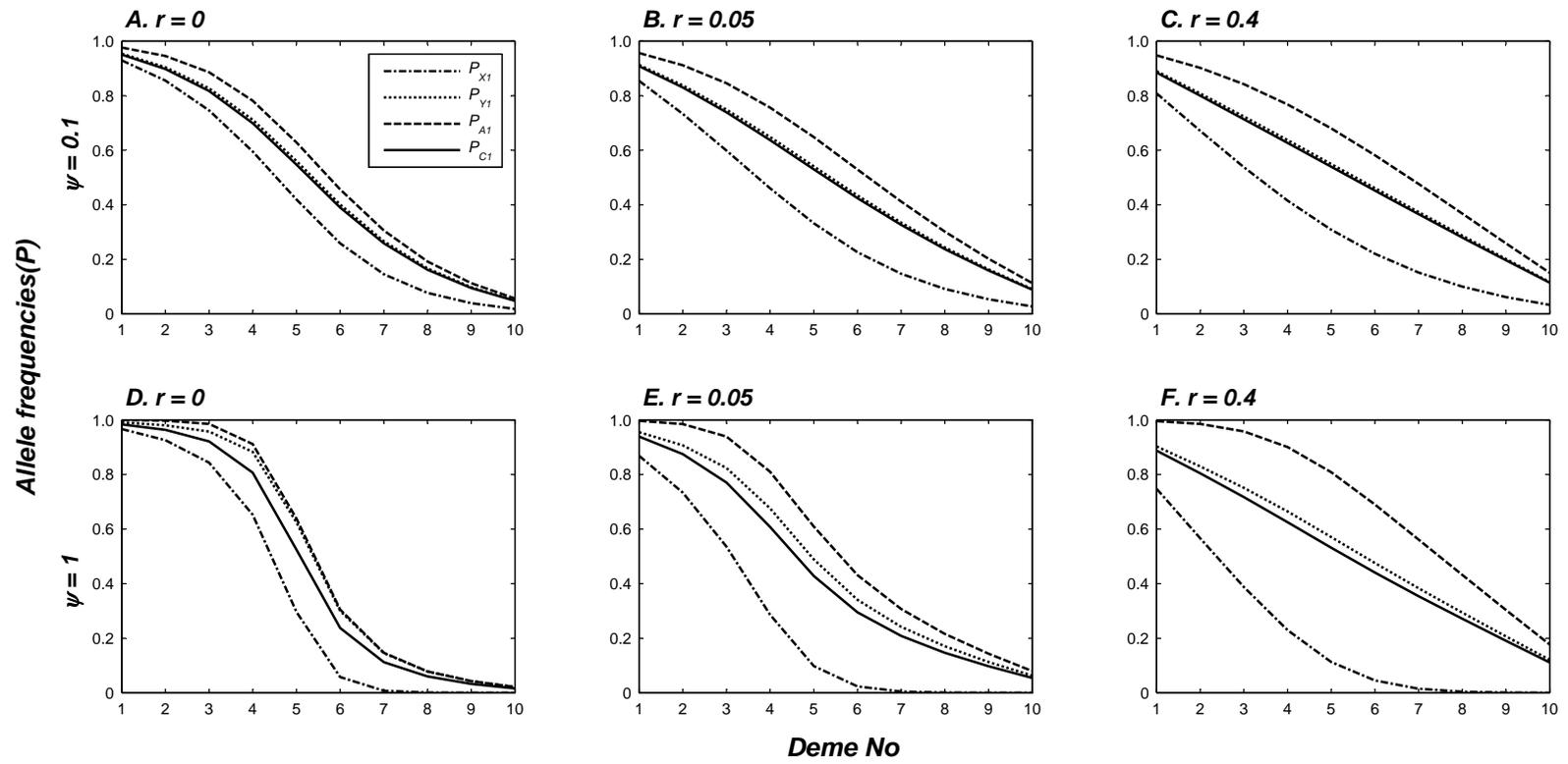

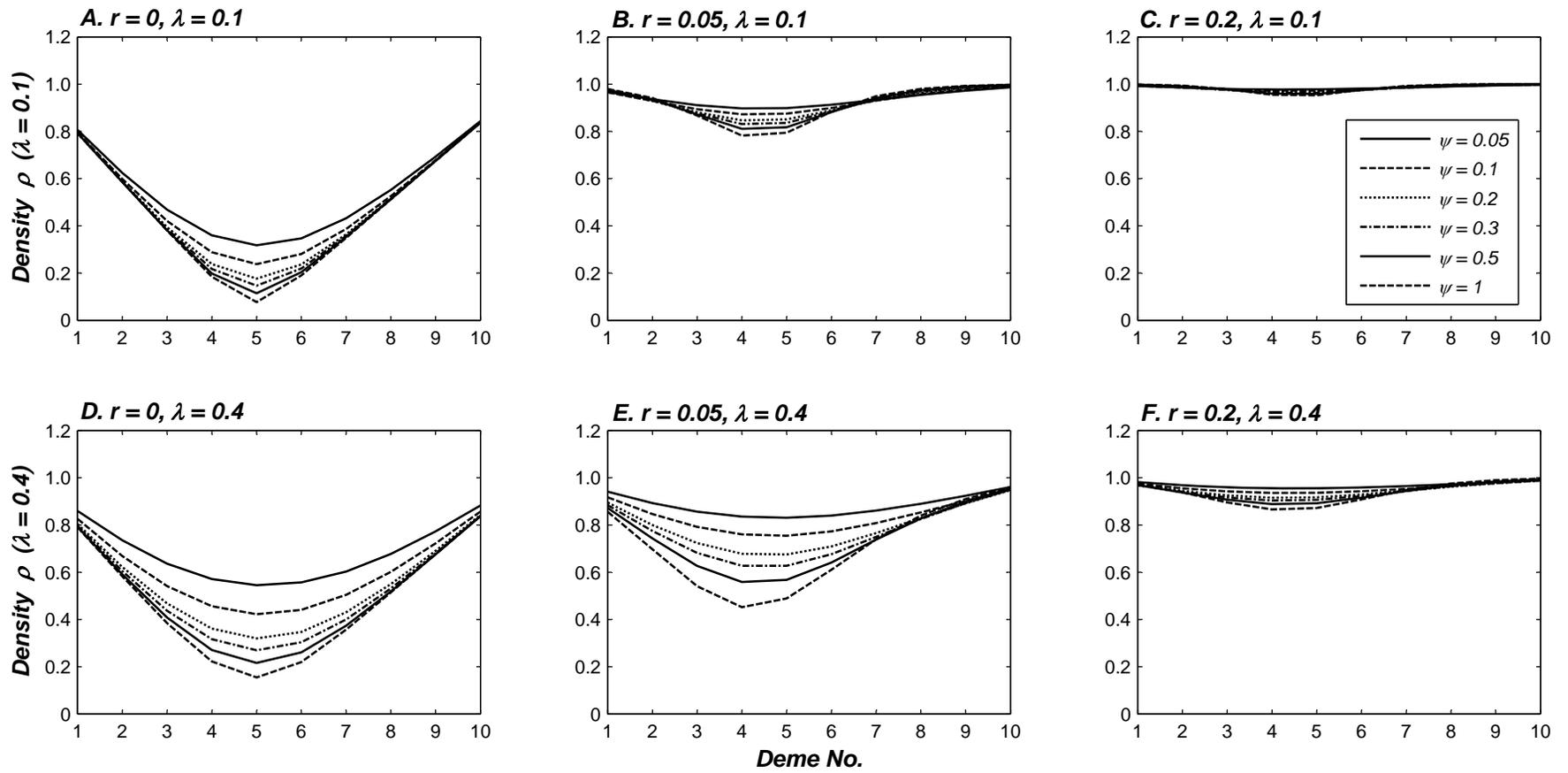

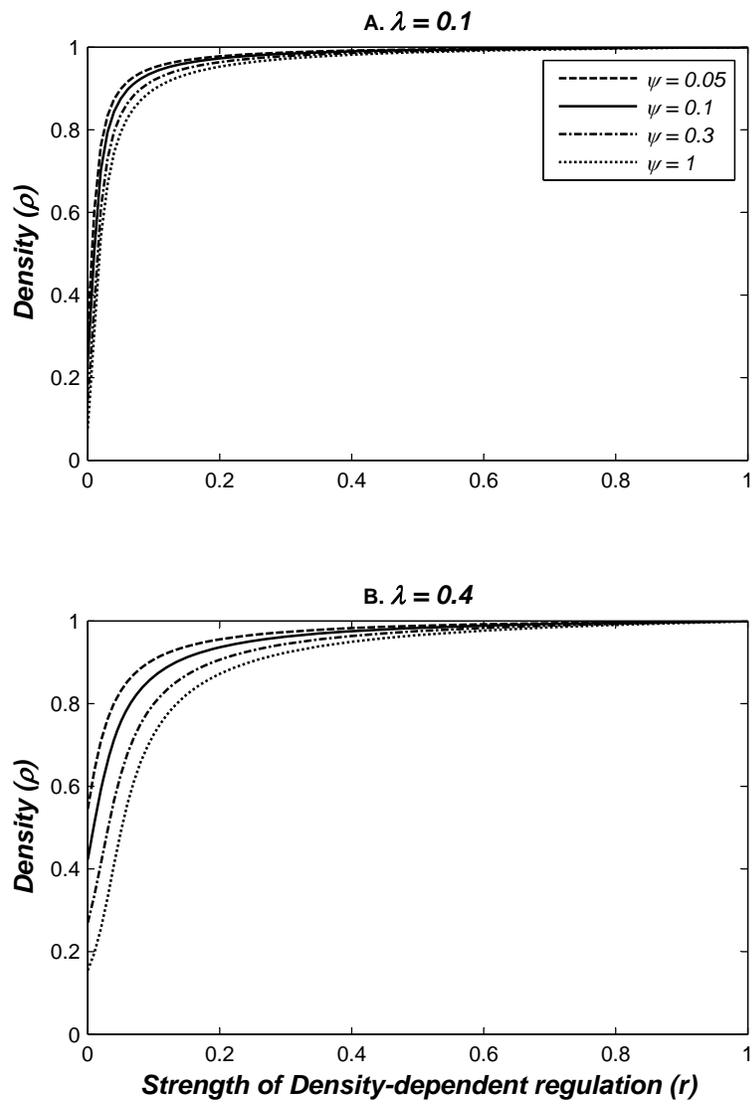

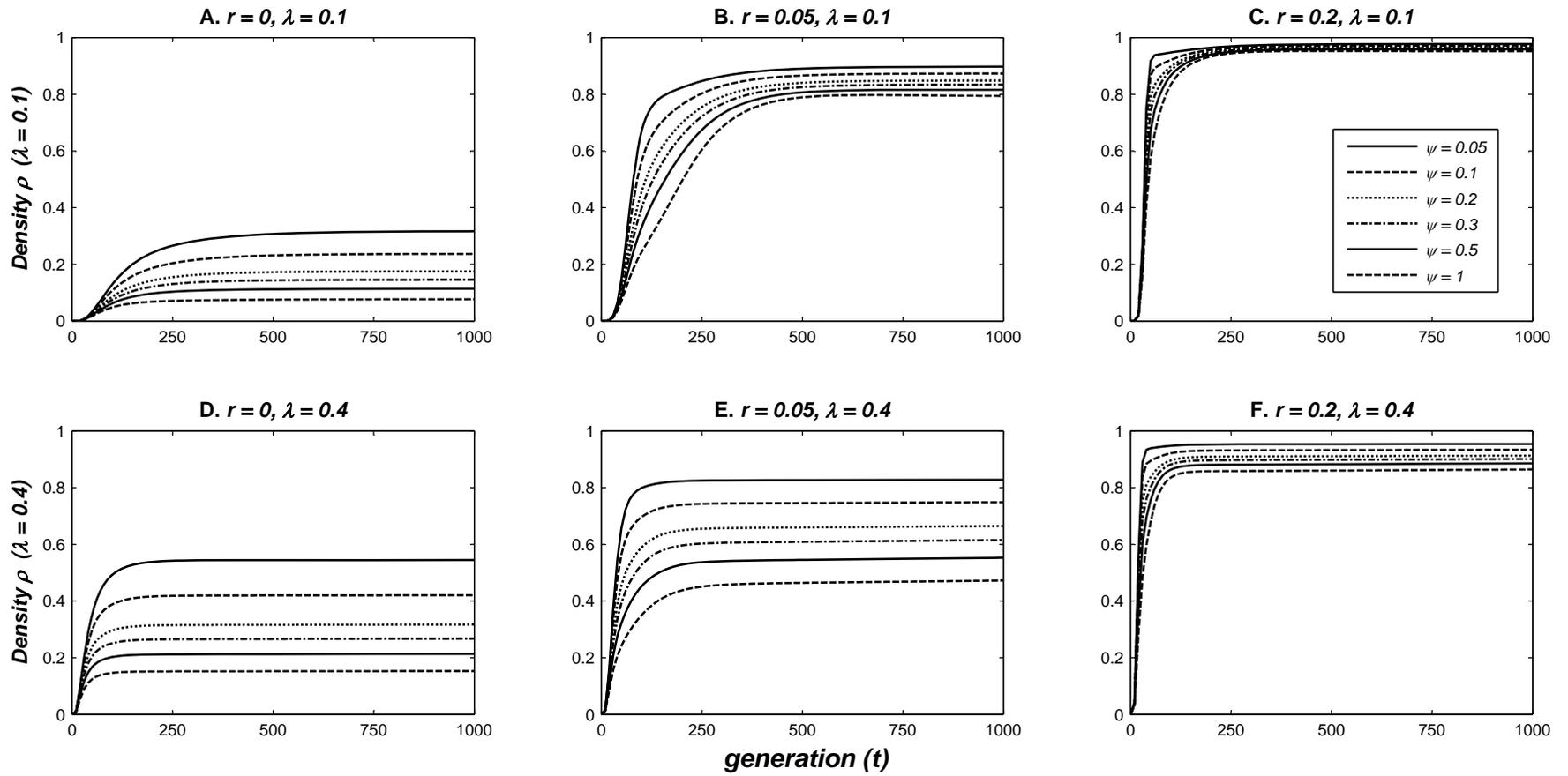

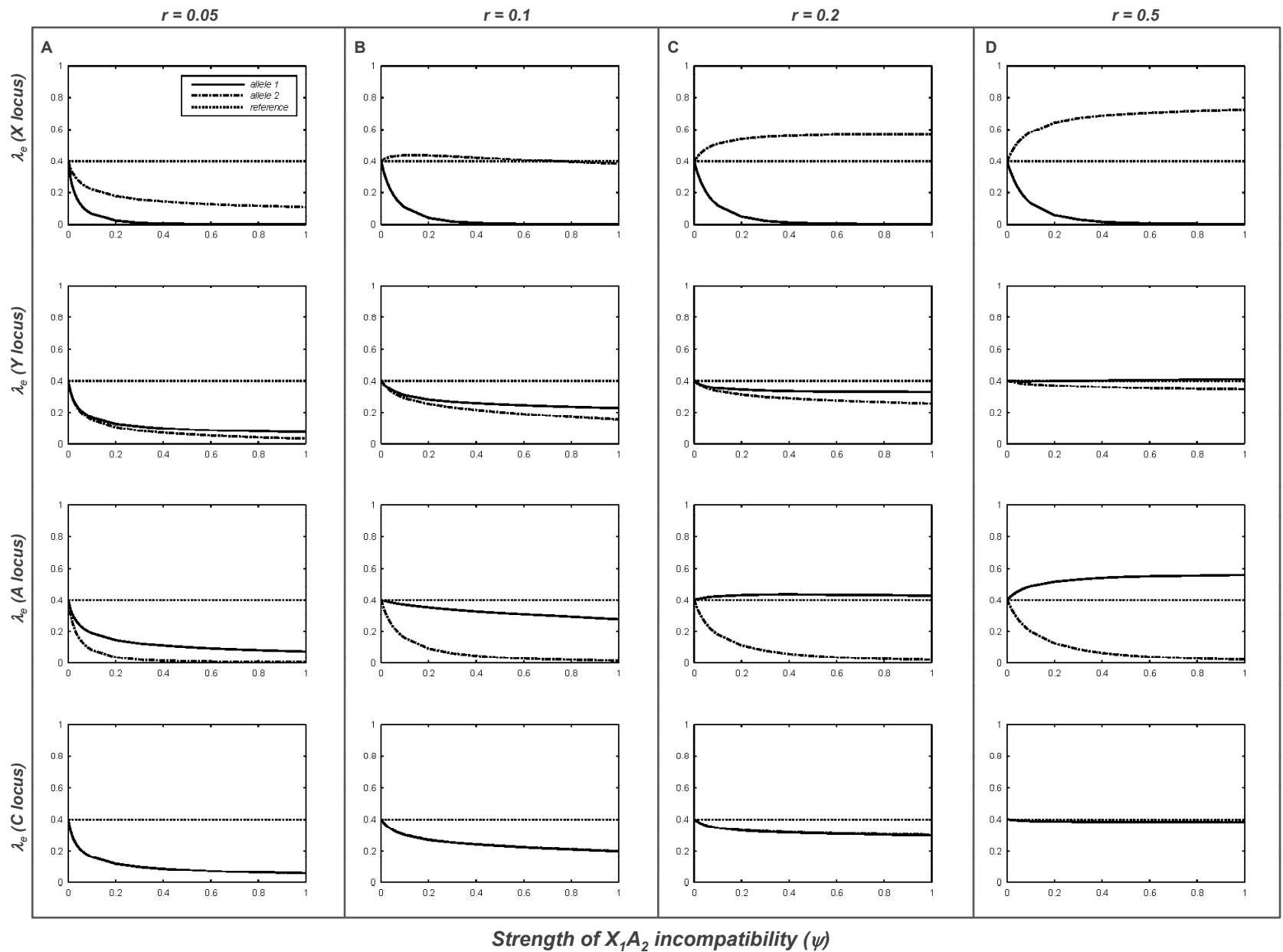

*Strength of $X_1A_2$ incompatibility ($\psi$)*

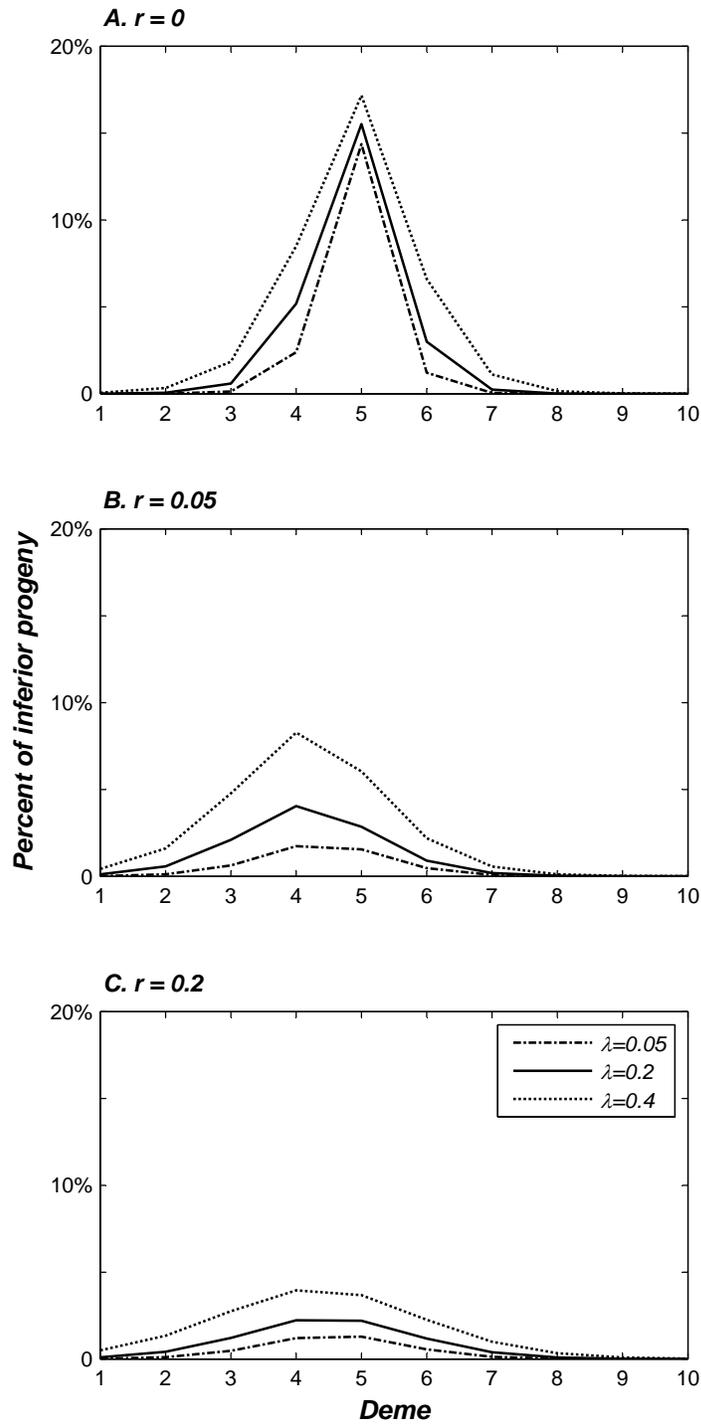

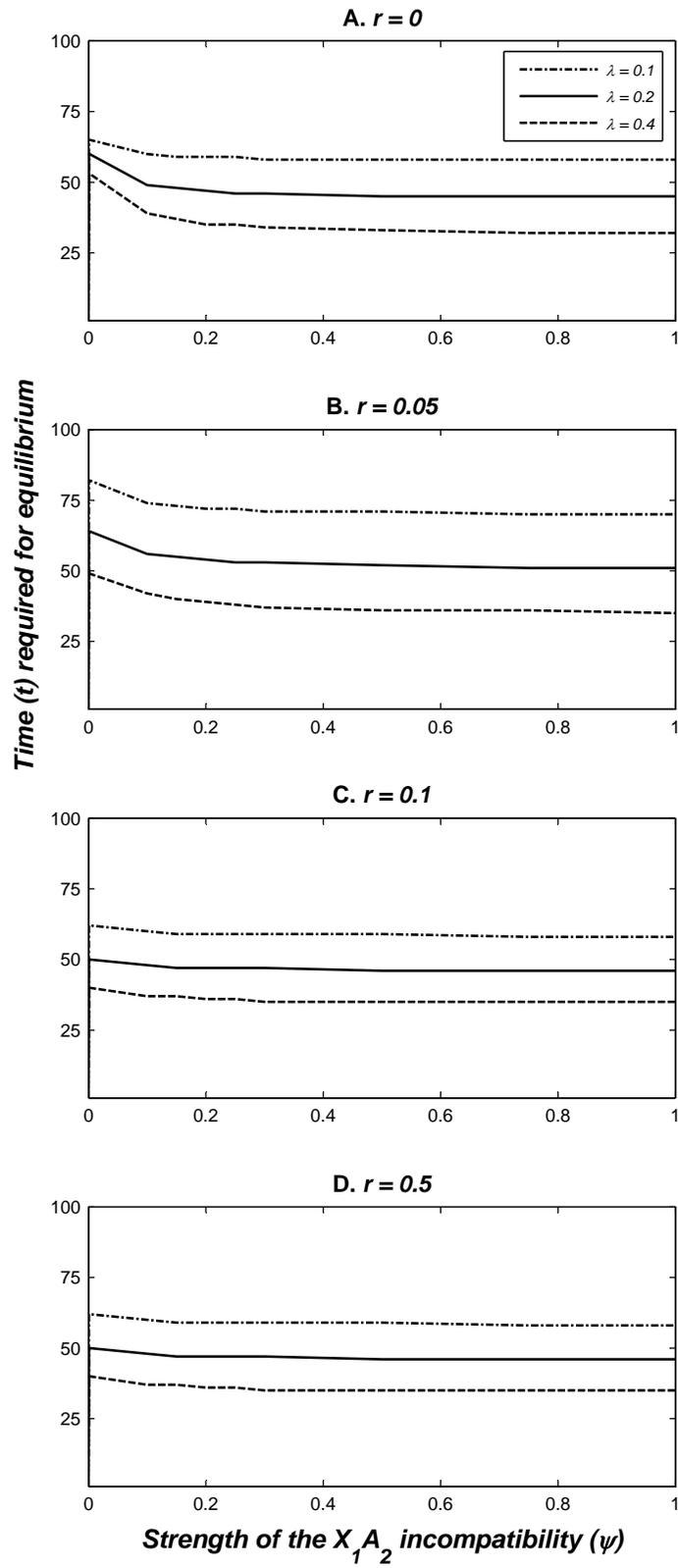